\documentclass[10pt,twocolumn,showpacs,amsmath,amssymb,aps,prl,floatfix]{revtex4-1}
\usepackage{amsmath}
\usepackage{graphicx}
\usepackage{amssymb}
\usepackage{color}

\newcommand{\comm}[1]{\ensuremath{\left[#1\right]_-}}

\newcommand{\fn}{\ensuremath{\mathbf{n}}}
\newcommand{\fs}{\ensuremath{\mathbf{s}}}
\newcommand{\fJ}{\ensuremath{\mathbf{J}}}
\newcommand{\dsd}{\ensuremath{\Delta_{\text{sd}}}}

\newcommand{\dra}{\ensuremath{\Delta_{\text{R}}}}

\newcommand{\fJp}{\ensuremath{\mathbf{J}_r}}
\newcommand{\fJph}{\ensuremath{\hat{\mathbf{J}}_r}}

\newcommand{\fJpr}{\ensuremath{\mathbf{J}_r}^{\rm{rel}}}

\newcommand{\fey}{\ensuremath{\mathbf e_{y}}}

\newcommand{\fmp}{\ensuremath{\mathbf m_r}}

\newcommand{\half}{\ensuremath{\frac{1}{2}}}

\newcommand{\mean}[1]{\ensuremath{\left\langle#1\right\rangle}}

\newcommand{\dx}{\ensuremath{\partial_x}}
\newcommand{\dt}{\ensuremath{\partial_t}}

\newcommand{\dpp}{\ensuremath{\partial_{(r)}}}

\newcommand{\ar}{\ensuremath{\alpha_{\text{R}}}}
\newcommand{\fmb}{\ensuremath{\overline{\mathbf m}_r}}
\newcommand{\dfmb}{\ensuremath{\overline{\partial_{(r)}\mathbf m}_r}}

\newcommand{\xio}{\ensuremath{x^{\infty}_{\rm{osc}}}}

\begin{document}
\title{Nonlocal spin torques in Rashba quantum wires with steep magnetic
textures}

\author{Martin Stier and Michael Thorwart}
\affiliation{I. Institut f\"ur Theoretische Physik, Universit\"at Hamburg, 
Jungiusstra{\ss}e 9, 20355 Hamburg, Germany}

\begin{abstract}
We provide a general procedure to calculate the current-induced spin-transfer
torque which acts on a general steep magnetic texture due to the exchange
interaction with 
an applied spin-polarized current. As an example, we consider a one-dimensional
ferromagnetic quantum wire and also include a Rashba spin-orbit interaction.
The spin-transfer torque becomes generally spatially non-local. Likewise, the
Rashba spin-orbit interaction induces a spatially nonlocal field-like
nonequilibrium spin-transfer torque. We also find a spatially varying
nonadiabaticity parameter and markedly different domain wall dynamics for very
steep textures as compared to wide domain walls. 
\end{abstract}


\maketitle

The exchange interaction of a spin-polarized electron current with localized
magnetic moments in a ferromagnetic wire typically induces a spin transfer
torque (STT). A pronounced consequence is the
coordinated switching of the localized magnetic moments of a ferromagnetic
domain wall (DW) in the wire generating a net DW motion 
\cite{beach2008current,obata2008current,parkin2008magnetic}. Other magnetic
textures also rose to recent prominence, such as magnetic vortices and
skyrmions \cite{yu2012skyrmion,romming2013writing,duine2014}, 
or one-dimensional spin chains\cite{khajetoorians2011}. 
There, the magnetization changes on much shorter length scales as 
compared to the conventional broad mesoscopic Bloch domain walls. In
addition, these textures are in general strongly affected by 
symmetry breaking interactions, such as the spin-orbit \cite{zhang2011skyrmions}
or the Dzyaloshinskii-Moriya interaction \cite{bode2007chiral,thiaville14}.\\
Clearly, in small structures, the backaction of
the local magnetic moments on the polarized itinerant electron spins 
can become relevant. In particular, they themselves experience a back-acting STT
as well. For wide magnetic textures, the impact of
backaction is generally small since the
itinerant spins relax much faster than they have time to 
interact with the local moments. Hence, for wide textures, it is reasonable to
assume 
a stationary spin polarized current which generates the (non-)adiabatic STT 
\cite{li2004}. This assumption, however, becomes increasingly invalid in
more narrow or steep magnetic textures. In this context, questions have been
raised why the nonadiabaticity parameter $\beta$ is much larger in small vortex
structures \cite{heyne2010,roessler2014} than compared to spin-waves 
in extended structures \cite{sekiguchi2012}. This effect has been
traced back to a non-standard description of the STT \cite{gonzalez2012}.\\
For a unified description of the STT for arbitrary magnetic textures,
several approaches have been developed 
\cite{waintal2004,xiao2006,bohlens2010,taniguchi2009,tatara2007,gonzalez2012,
nguyen2007,wang2012}. Nevertheless, a complete picture is still missing.
Spin relaxation has either been neglected \cite{waintal2004,xiao2006} 
or included \cite{bohlens2010,taniguchi2009,tatara2007}, quantum corrections are
considered\cite{ohe2006} or a semiclassical
approach on the basis of spin diffusion has been formulated \cite{gonzalez2012,akosa2015}.
Spin-orbit interaction has been considered for broad textures
\cite{nguyen2007,wang2012} only.\\
In this work, we provide a general and conceptually simple scheme to calculate
the STT for arbitrary magnetic textures. To show the flexibility
 of the approach, 
we also include the Rashba spin-orbit interaction in the itinerant
electrons, whose contribution to the STT in steep magnetic structures is of 
high interest \cite{brataas2014spin,brataas2014sot}. The STT and the
resulting nonadiabatic Rashba field-like STT are shown to become non-local in
space for steep textures. The known results for broad Bloch DWs
\cite{thorwart2007,stier2013,stier2014} are recovered as a limiting case. 
We calculate DW velocities for a broad range of widths of a Bloch DW and find
a non-local nonadiabaticity parameter for steep textures. Interestingly, the
overall direction of the DW movement can be determined by the DW width.\\
We consider a one-dimensional (1D) ferromagnetic quantum wire 
with a magnetic texture formed by localized magnetic moments $M_s\fn(x,t)$ with 
a saturation magnetization $M_s$ and unit vectors $\fn(x,t)$. Their dynamics
follows from the Landau-Lifshitz-Gilbert (LLG) equation  
\begin{equation}
 \dt\fn = -\gamma_0\fn\times\mathbf H_{\rm{eff}} + \alpha \fn\times\dt\fn
+\mathbf T \label{eq::llg}\ ,
\end{equation}
with the effective magnetic field $\mathbf H_{\rm{eff}}$, the gyro-magnetic
ratio $\gamma_0$, and the Gilbert damping constant $\alpha$. $\mathbf T$
denotes the spin torque which is induced by the exchange interaction of the
localized moments with the polarized spins of the current carrying
electrons. The latter and the interaction are described by the 
Hamiltonian
\begin{equation}
 H = H_{\rm{kin}}  + H_{\rm{Rashba}} + H_{\rm{sd}} + H_{\rm{relax}}\, 
. \label{eq::totham}
\end{equation}
It contains the kinetic energy, the Rashba spin-orbit interaction, the
$sd$ interaction with the magnetic moments and the relaxation of
the electron spins. It is convenient to use the low-energy description of
this Hamiltonian, as it is accurate for 1D quantum wires 
\cite{gogolin2004bosonization} and yields a simple derivation of the
STT \cite{thorwart2007,stier2013}. Then, all relevant
electron operators $c_{\sigma},c^{\dagger}_{\sigma}$ with spin index $\sigma$
can be included in spin-like operators
$\fJph=\half:c^{\dagger}_{r\sigma}\boldsymbol\sigma_{\sigma\sigma'}
c_{r\sigma'}:$. The index $r=R/L=+/-$ refers to right or left moving electrons 
in the wire. The Rashba Hamiltonian assumes the simple form
$H_{\rm{Rashba}}= \dra\sum_{r} p\int dx\ \fJph\cdot \mathbf e_y$, with the Rashba
splitting $\dra=2\tilde\alpha_{\text{R}} k_F$\cite{stier2014}. All terms of
Eq.\ (\ref{eq::totham}) can be expressed in terms of $\fJph$, for which the
Heisenberg equations of motion $\dt\fJph=-\frac{i}{\hbar}\comm{\fJph,H}$
can readily be evaluated \cite{thorwart2007, stier2013, stier2014}. 
We find for the expectation value $\fJp =\langle\fJph\rangle$
\begin{align}
 \dpp\fJp
=&-\frac{\dsd}{\hbar}[\fJp\times\fmp+\beta(\fJp-\fJp^{\rm{rel}})]\ ,
\label{eq::eom}
\end{align}
with the derivative $\dpp=\dt+vr\dx$, $\fmp=\fn+r\ar\fey$, the
Fermi velocity $v$, the exchange interaction strength 
$\dsd$, the Rashba parameter $\ar=\dra/\dsd$, the nonadiabaticity
parameter $\beta=\hbar/(\dsd\tau)$, relaxation time $\tau$ and the relaxed spin
density $\fJp^{\rm{rel}}$. In fact, this is the continuity equation
of the spin current\cite{zhou2009spin,sun2005,zhang2004roles,shi2006} when we introduce the spin density
$\fs=\fJ_R+\fJ_L$ and the spin current density $\fJ=v(\fJ_R-\fJ_L)$ and sum over
$r$.\\
To solve the equations of motion (\ref{eq::eom}), we set up the ansatz 
\begin{equation}
 \fJp=a_r\fmb+b_r\dfmb+c_r\fmb\times\dfmb\ ,\label{eq::ansatz}
\end{equation}
 with the unit vectors $\fmb=\fmp/|\fmp|$ and $\dfmb=\dpp\fmb/|\dpp\fmb|$. It is
the central observation of this work to use space- and time-dependent
coefficients $a_r(x,t),b_r(x,t),c_r(x,t)$. All three vectors in Eq.\
(\ref{eq::ansatz}) form a complete
orthonormal basis in spin space. With the prefactors determined below, 
the STT $\mathbf T =
-\frac{\dsd}{\hbar}\fn\times\fs=-\frac{\dsd}{\hbar}\fn\times\sum_r\fJp$ can
be calculated. Straightforward algebra yields 
\begin{align}
 \mathbf T = \sum_{\nu=x,t}&
\left[T_{\nu}^{\rm{nonad}}(b_r)\overline{\partial_{\nu}\fn}
+T_{\nu}^{\rm{ad}}(c_r)\fn\times\overline{\partial_{\nu}\fn}\right] \nonumber 
\\
&-H_{\text{R}}(a_r,b_r)\fn\times\fey \, . \label{eq::stt}
\end{align}
 In particular, the ansatz yields the adiabatic STT 
 $T_{x,t}^{\rm ad}$, the nonadiabatic STT $T_{x,t}^{\rm nonad}$, as well
as the field-like Rashba term with the nonadiabatic Rashba field $H_{\text{R}}$. By
construction, the antidamping field $\mathbf H_{\text{R}}^{\rm
anti}=-\beta\fn\times\mathbf H_{\text{R}}$ is included, 
although it does not appear explicitly. It easily  can be recovered by
choosing a suitable overcomplete basis in the ansatz
(\ref{eq::ansatz}). Notice that we have defined the STT in terms of the 
normalized vectors of the derivative $\overline{\dx\fn} =
\dx\fn/|\dx\fn|$ to avoid a divergence of the prefactors. In general, also terms
$\propto\dt\fn,\fn\times\dt\fn$ appear. They act as additional damping terms in
Eq.\ (\ref{eq::llg}) and are conveniently absorbed in a rescaled Gilbert
parameter.\\
The remaining step is to determine the coefficients $a_r,b_r$ and $c_r$. For
this, we insert the Ansatz Eq.\  (\ref{eq::ansatz}) 
into Eq.\ (\ref{eq::eom}) and order it according to linearly independent
parts. This yields to an ordinary differential equation 
\begin{equation}
 \dpp \begin{pmatrix}a_r\\b_r\\c_r\end{pmatrix} = A_r
\begin{pmatrix}a_r\\b_r\\c_r\end{pmatrix} 
 + \frac{\dsd\beta}{\hbar}\begin{pmatrix}a_r^{\text{rel}}\\ b_r^{\text{rel}} \\
c_r^{\text{rel}}
\end{pmatrix},\label{eq::dgl}
\end{equation}
where $a_r^{\text{rel}}$, $b_r^{\text{rel}}$ and $c_r^{\text{rel}}$ are the
corresponding coefficients of the relaxed spin
density $\fJp^{\rm{rel}}$. The coefficient matrix 
\begin{align}
 A_r = \begin{pmatrix}
-\dsd\beta/\hbar & |\dpp\fmb| & 0\\
-|\dpp\fmb| & -\dsd\beta/\hbar & -\dsd|\fmp|/\hbar-g\\
0 & \dsd|\fmp|/\hbar+g & -\dsd\beta/\hbar
            \end{pmatrix}\label{eq::coeff}
\end{align}
is space- and time-dependent due to $|\dpp\fmb|(x,t)$ and
$g=[\dpp\fmb\cdot(\fmb\times\dpp^2\fmb)]/(\dpp\fmb)^2$. Hence, 
a general analytical solution of Eq.\ (\ref{eq::dgl}) cannot be found. 
Even though these equations are
derived within a 1D model, they readily can be used for higher dimensional 
magnetic textures, if the electronic motion perpendicular to the
current is less important.\\
In the following, we solve Eq.\ (\ref{eq::dgl}) numerically for the 
example of a Bloch DW. Other textures can be treated in
the similar way.  As both Eqs.\ (\ref{eq::dgl}) and (\ref{eq::llg}) 
depend on $\fn$, it is necessary to solve them self-consistently for each 
time step until convergence is achieved. This is still
demanding and we can use physical arguments to
further simplify the equation. First, we may assume that all coefficients depend
only on the distance  to the DW center
$x_{\rm{DW}}(t)$, since we do not consider explicit time-dependent modifications
of the shape of the initially prepared domain wall \cite{thorwart2007}. Hence, 
$a_r(x,t)=a_r[\Delta x = x-x_{\rm{DW}}(t)]$. Then, it follows that 
$\dpp a_r(\Delta x) = [pv-\dt x_{\rm{DW}}(t)]\partial_{\Delta x}a_r(\Delta x)$.
As the DW velocity $\dt x_{\rm{DW}}(t)$ is generally 
much smaller than the Fermi velocity $v$, we can neglect the terms involving 
$\dt x_{\rm{DW}}(t)$. Effectively, every time derivative
in Eq.\ (\ref{eq::dgl}) is neglected, and thereby also the damping parts in Eq.
(\ref{eq::stt}). We do not simplify the LLG equation (\ref{eq::llg}) by this
assumption, so that a precessional motion of the DW is still possible.\\
\begin{figure}[tb]
\centering
\includegraphics[width = \linewidth]{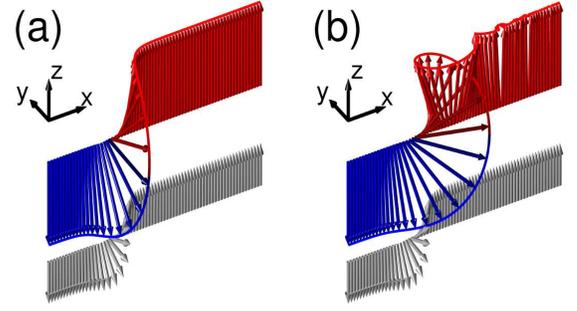}
 \caption{\label{fig::scheme}(Color online) Sketch of the configurations of the
localized 
DW spins  (gray arrows) and the itinerant spins flowing from negative to
positive $x$-values, for a broad (a) and a steep 
(b) domain wall 
(distances are normalized to the DW width). Red (blue) arrows show the direction
of the 
$z$-component of the itinerant spins in positive (negative) direction. 
The lines are guides for the eyes. While for broad DWs ($\lambda\gg x_{\rm osc}^{\infty}$), the itinerant
spins 
follow the DW magnetization adiabatically, this is not possible for steep DWs ($\lambda\leq x_{\rm osc}^{\infty}$).}
\end{figure}
Before we discuss the numerical results, it is instructive to analyze the
generic qualitative behavior which can be extracted from Eq.\ (\ref{eq::dgl}). 
The eigenvalues of the matrix in Eq.\ (\ref{eq::coeff}) are of the form 
$\xi_1=-\beta\dsd/(\hbar v)$ and $\xi_{\pm}=[-\beta\dsd\pm
i\sqrt{(\dsd|\fmp|+g)^2+ (\hbar v \dx|\fmp|)^2} ]/(\hbar v)$. 
For constant coefficients, the solutions of
Eq.\ (\ref{eq::dgl}) would be combinations of exponentials 
$\propto \exp(\xi_i x)$. Thus, the real part $\text{Re}\, \xi_i$ determines 
a damped, and the imaginary part
$\text{Im} \xi_i$ an oscillating behavior. As the real part is the same for
all $\xi_i$, we find a unique damping length
$x_{\rm{damp}}= \hbar v / (\beta\dsd)$. The oscillation length $2\pi
x_{\rm{osc}}$ can only be estimated. 
Far away from the DW center, when 
$|\dx\fmb|\propto|\dx\fn|=0$, it is determined by the
inverse \textit{sd} coupling strength such that $x_{\rm{osc}}\approx
\xio \equiv  \hbar v/\dsd$. 
Close to the DW center, $x_{\rm{osc}}$ is modified by the term $|\dx\fmb| 
\propto 1/\lambda$, where $\lambda$ is the DW width. Thus, in particular for
steep DWs, the oscillation period is reduced.\\
\begin{figure}[tb]
\centering
 \includegraphics[width=.8\linewidth]{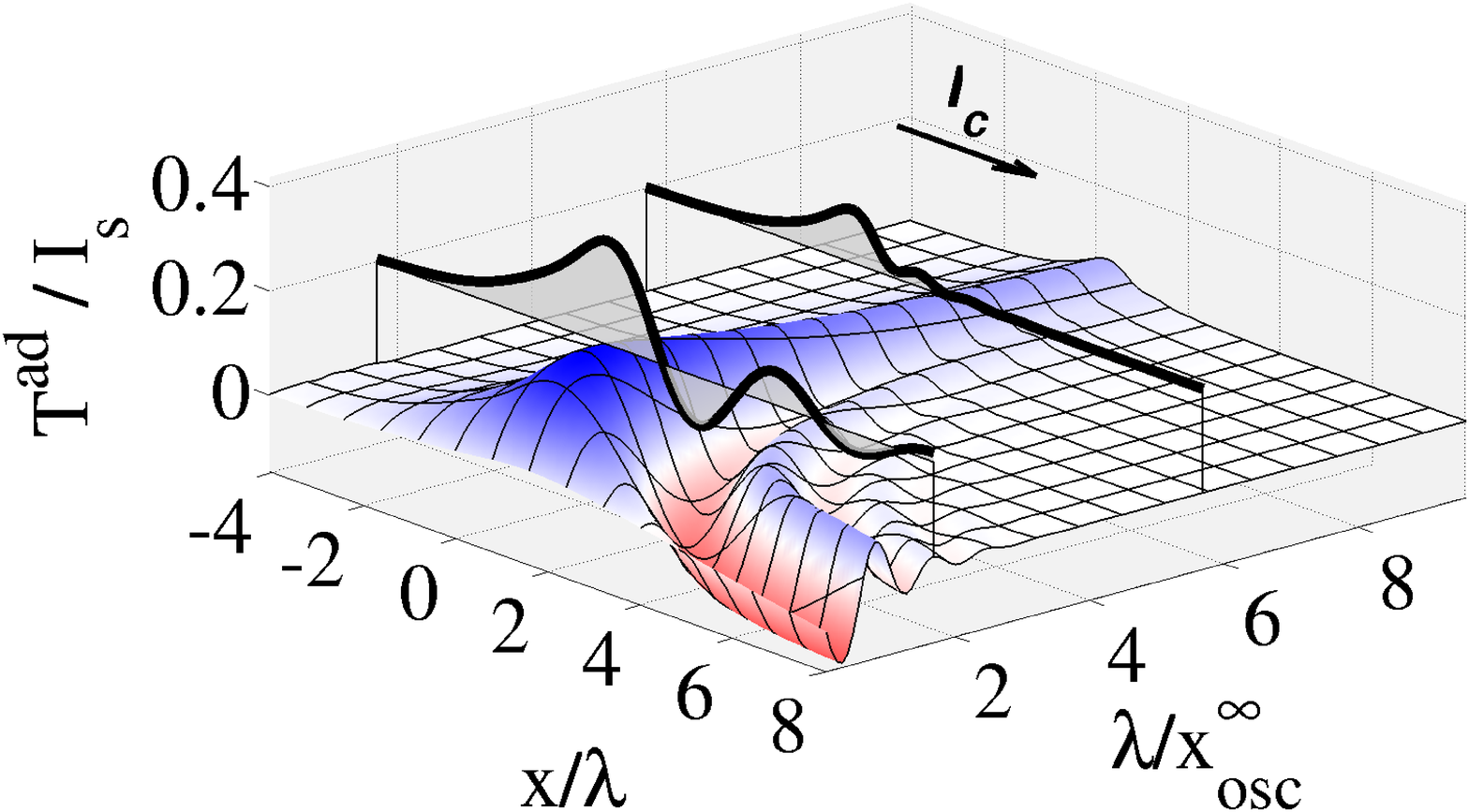}
 \includegraphics[width=.8\linewidth]{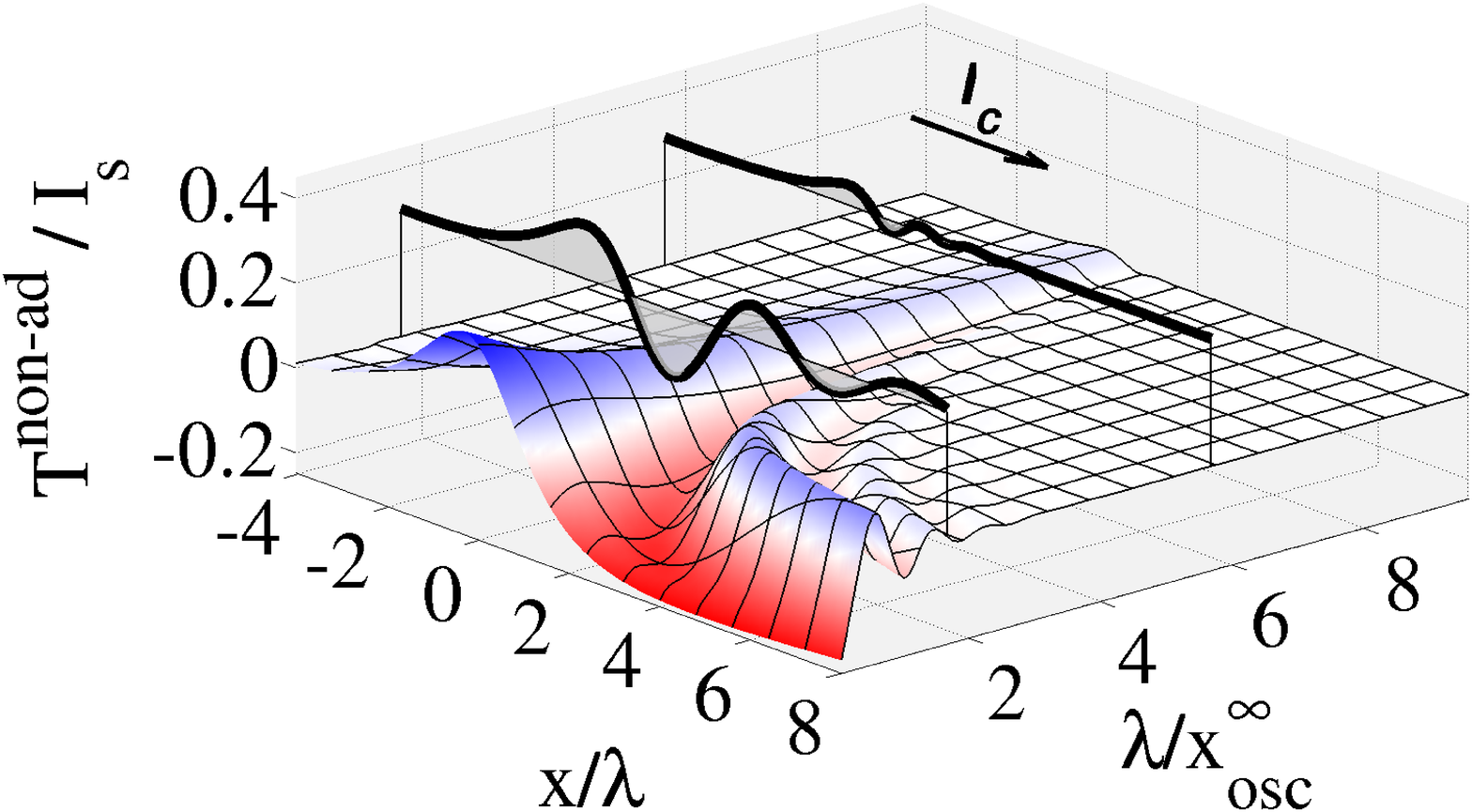}
 \includegraphics[width=.8\linewidth]{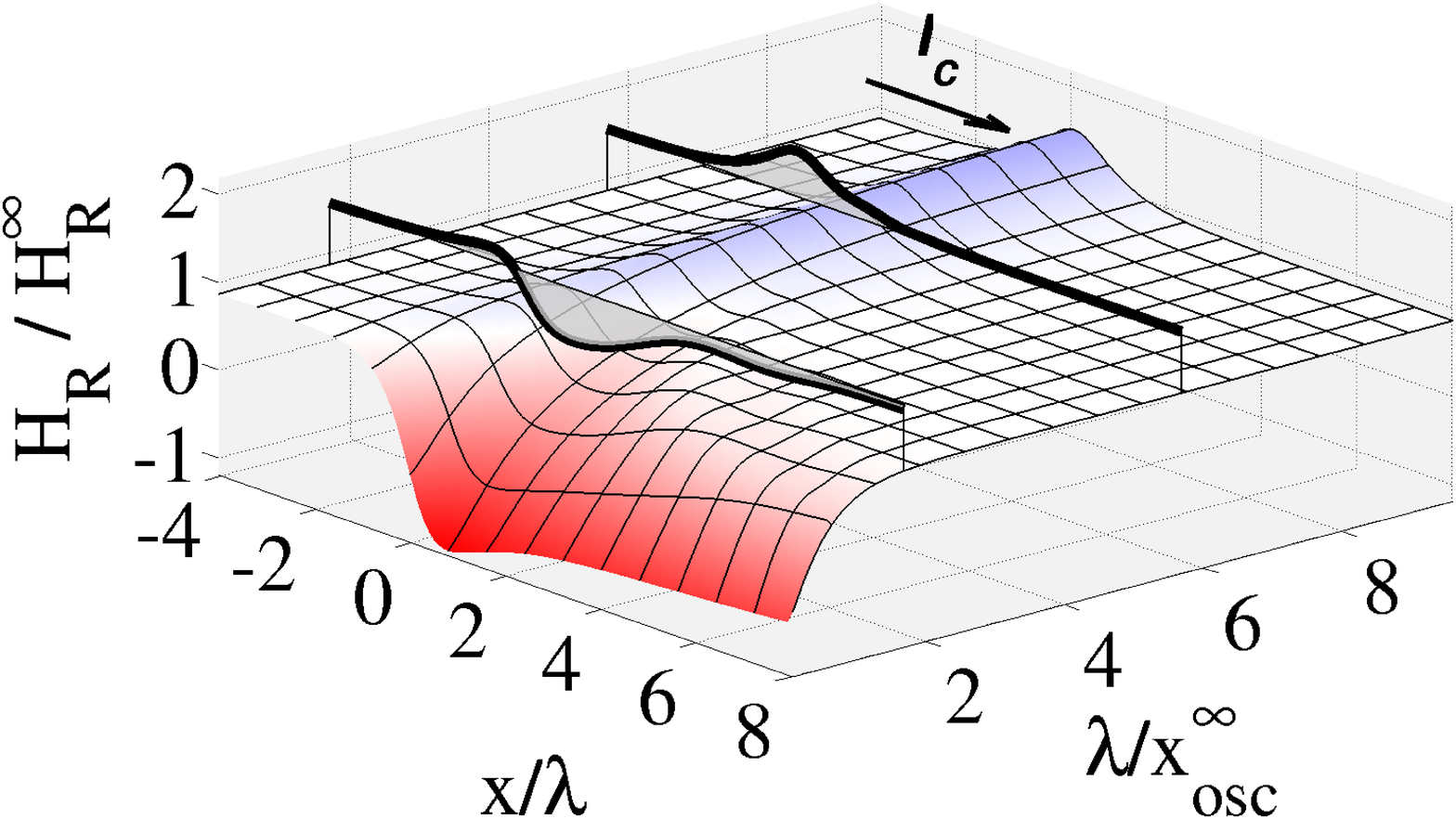}
 \caption{\label{fig::stt}(Color online) Spatial dependence of the 
three components to the STT: the adiabatic STT (top), the nonadiabatic STT
(middle), 
 and, the Rashba field-like term (bottom) for varying 
 DW widths $\lambda$. Blue (red) colors indicate positive (negative) deviation
from their
relaxed values. Two cases for $\lambda/x^{\infty}_{\rm{osc}}=1.7,5.7$ are
highlighted (shifted thick black curves). 
The parameters are $\ar=0.4$, $\beta=0.2$. }
\end{figure}
The qualitative observations are supported by explicit numerical calculations.
We 
consider a Bloch-z DW with boundary conditions 
$n_z(x\to\pm\infty)=\pm1$. The DW is formed by the effective field
$\mathbf H_{\rm{eff}}=(2A_{\rm{ex}}/M_s)\dx^2\fn +K_{\parallel} n_{\parallel}
\mathbf e_{\parallel}-K_{\perp} n_{\perp}  \mathbf e_{\perp}$, 
with an easy (hard) axis along $\mathbf e_{\parallel(\perp)}$. Further, we choose the physical
parameters of Pt/Co/AlO$_x$ and set $A_{\rm ex}=10^{-11}\rm{J/m}$ 
and $M_s=1090\rm{kA/m}$. In addition, we fix $\alpha=0.1$ and 
set $\dsd=0.25\rm{eV}$, $v_F=10^6\rm{m/s}$ and the current's polarization $P=1$. The Rashba interaction is chosen 
around $\tilde\ar=10^{-10}\rm{eVm}$ corresponding to
$\Delta_{\text{R}}=0.1\rm{eV}$
and $\ar=\Delta_{\text{R}}/\dsd=0.4$. We vary the DW width
$\lambda=\sqrt{J/K_{\parallel}}$, with $J=2A_{\rm ex}/M_s$, by
setting $J = \lambda J^{(0)}$ and $K_{\parallel} =
K_{\parallel}^{(0)}/\lambda$ for a fixed ratio 
$\sqrt{J^{(0)}/K^{(0)}_{\parallel}}=1$. 
The perpendicular anisotropy is set to $K_{\perp}=0.3 K_{\parallel}$. 
We inject polarized electrons from the right in their relaxed state. Thus,   
$\fJ_R(x\to-\infty)=\fJ_R^{\rm{rel}}$, with $\fJ_R^{\rm{rel}}=(I_s/v)\fn$ 
with the saturation spin current $I_s\equiv P I_c /(2eM_s)$. Moreover, we use 
the oscillation period 
$\xio=\hbar v / \dsd =2.6$ nm as a length scale. This uniquely determines 
the initial values of Eq.\ (\ref{eq::dgl}).\\
Two typical situations emerge and are 
illustrated in Fig. \ref{fig::scheme}. For broad DWs, the itinerant 
spins follow the DW adiabatically. In contrast, for steep DWs, the
spins experience a mismatch of $\fn$ and $\fJ_R$ particularly at and beyond the
DW center. 
This induces a significant backaction on the STT. 
In Fig.\ \ref{fig::stt}, the three contributions to ${\bf T}$ are shown  for
varying DW widths. For broad DWs ($\lambda \gg \xio$), 
the conventional contributions $T^{\rm ad}=I_s
|\dx\fn|$, $T^{\rm nonad}=-\beta I_s|\dx\fn|$ and 
$H^{\infty}_{\text{R}}=\dsd\ar I_s /(v\hbar \gamma_0)$ are recovered (cf. Appendix). 
In fact, Eq.\ (\ref{eq::dgl}) contains these solutions
for $\lambda\to\infty$. 
Differences occur for the Rashba field-like torque due to
corrections in first order in the derivative of the
magnetization \cite{stier2013,stier2014}. In contrast, 
for steep DWs ($\lambda < \xio$), every STT component is altered. 
 In particular, both the adiabatic and the nonadiabatic STT show spatial
oscillations which are damped after crossing the DW with the 
damping length $x_{\rm{damp}}= \hbar v / (\beta\dsd) = \beta^{-1}
x^{\infty}_{\rm{osc}}$. For sharp DWs 
($\lambda\ll x_{\rm{osc}}^{\infty}$), the adiabatic STT is suppressed while
the nonadiabatic STT is strongly increased. This behavior is even more
pronounced when we exclude relaxation ($\beta=0$). Then,
the oscillations remain undamped beyond the DW (cf. Appendix). Besides
the (non)adiabatic STT, the  Rashba field-like torque is also enhanced and may
even change its sign at the DW center. As we solved Eq.\ (\ref{eq::dgl}) exactly
no divergent terms appear for $\mathbf H_{\text{R}}$ which could be the case in
approximate calculations\cite{stier2014}.\\
\begin{figure}[tb]
\centering
\begin{minipage}{.49\linewidth}
$\beta=0$\\
 \includegraphics[width=.99\linewidth]{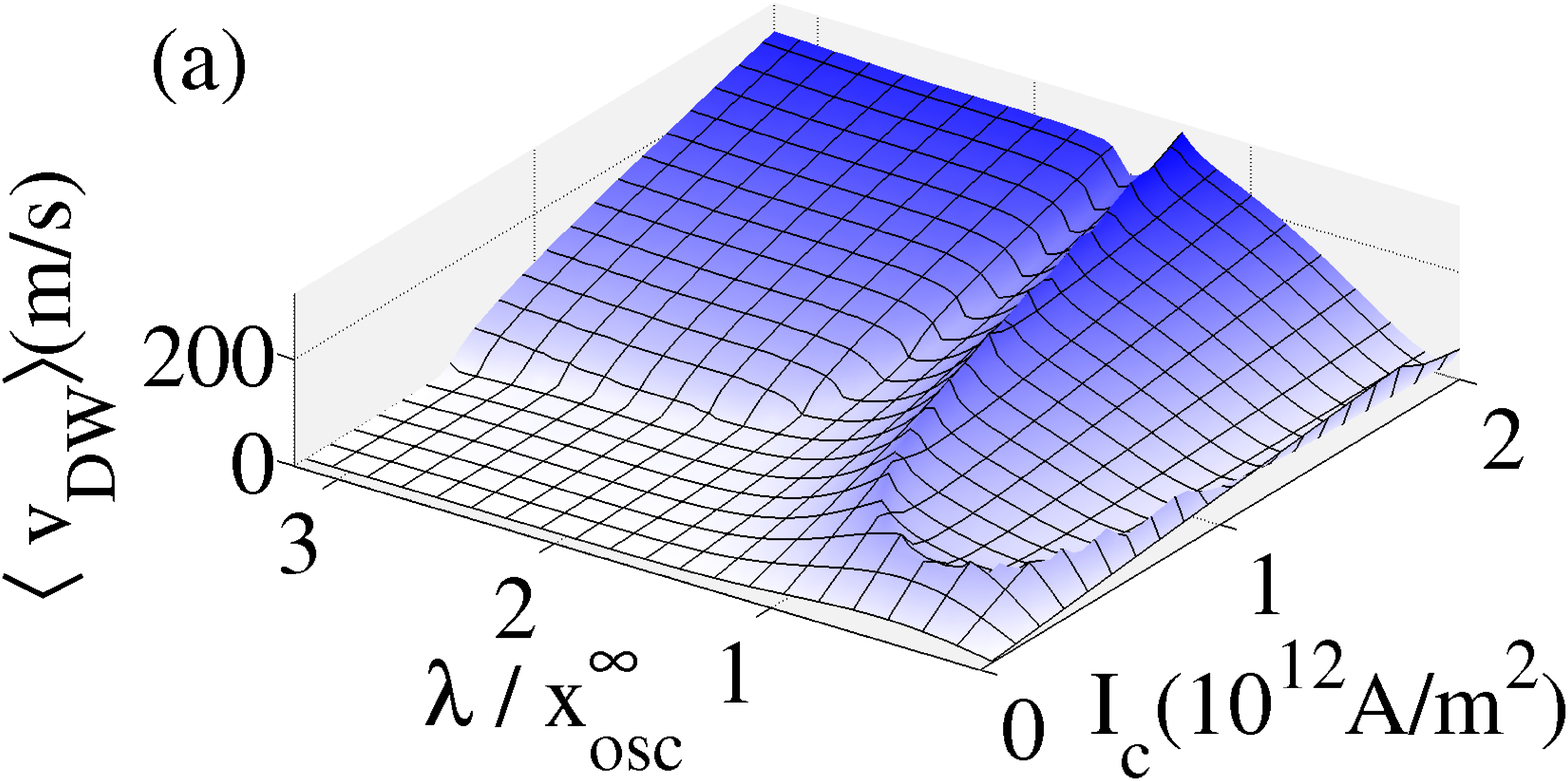} 
 \includegraphics[width=.99\linewidth]{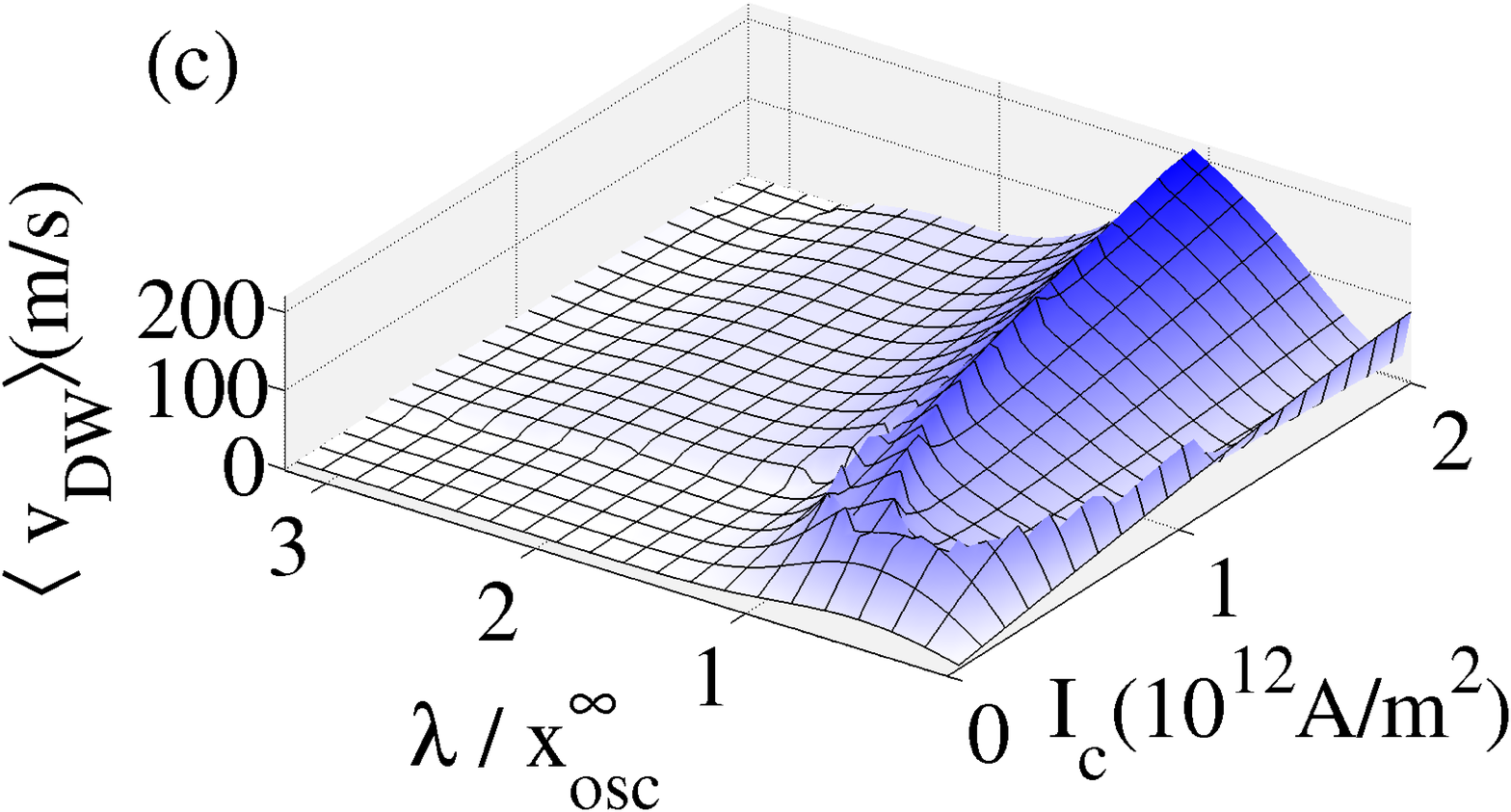} 
 \includegraphics[width=.99\linewidth]{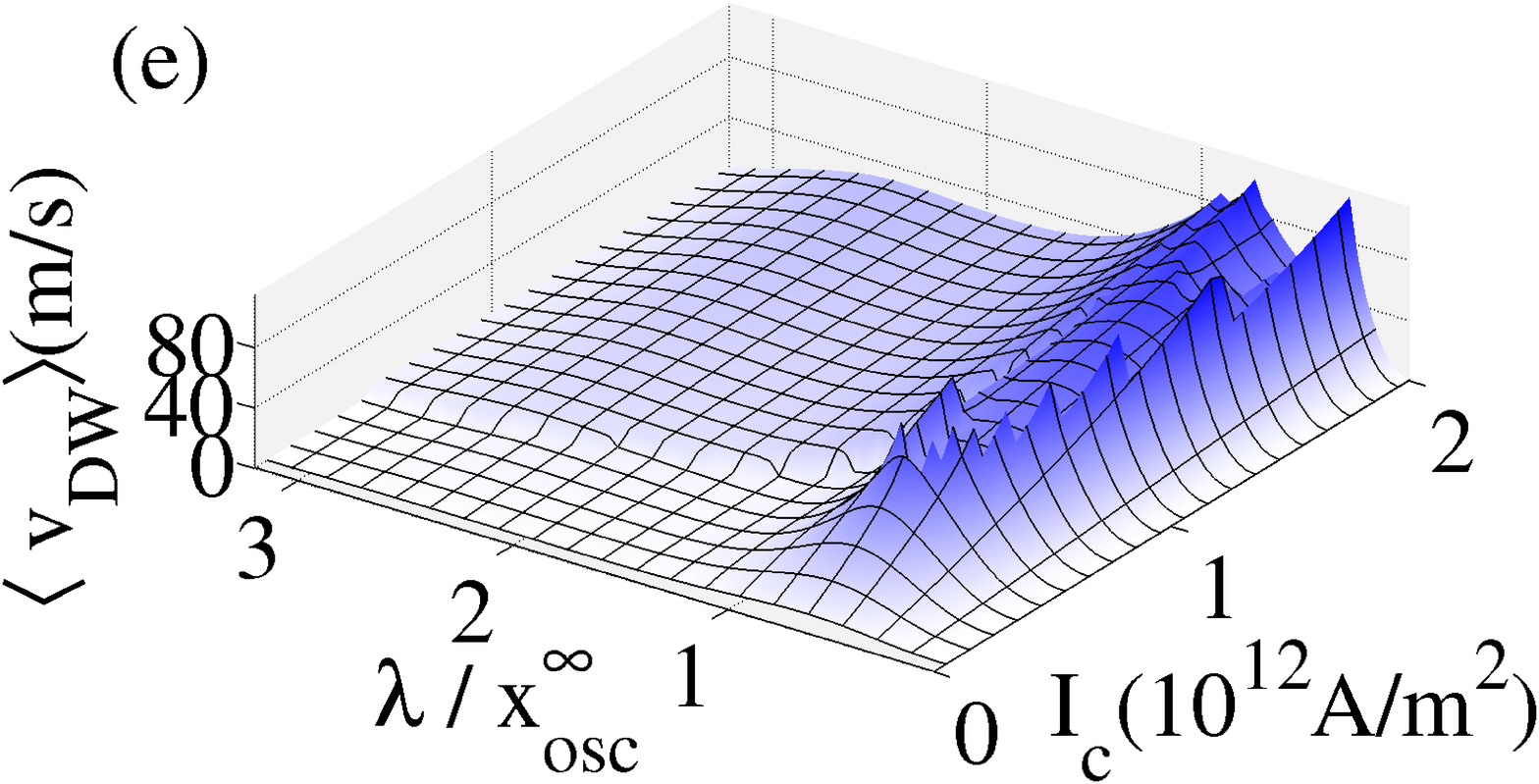} 
\end{minipage}
\begin{minipage}{.49\linewidth}
$\beta=0.2$\\
\includegraphics[width=.99\linewidth]{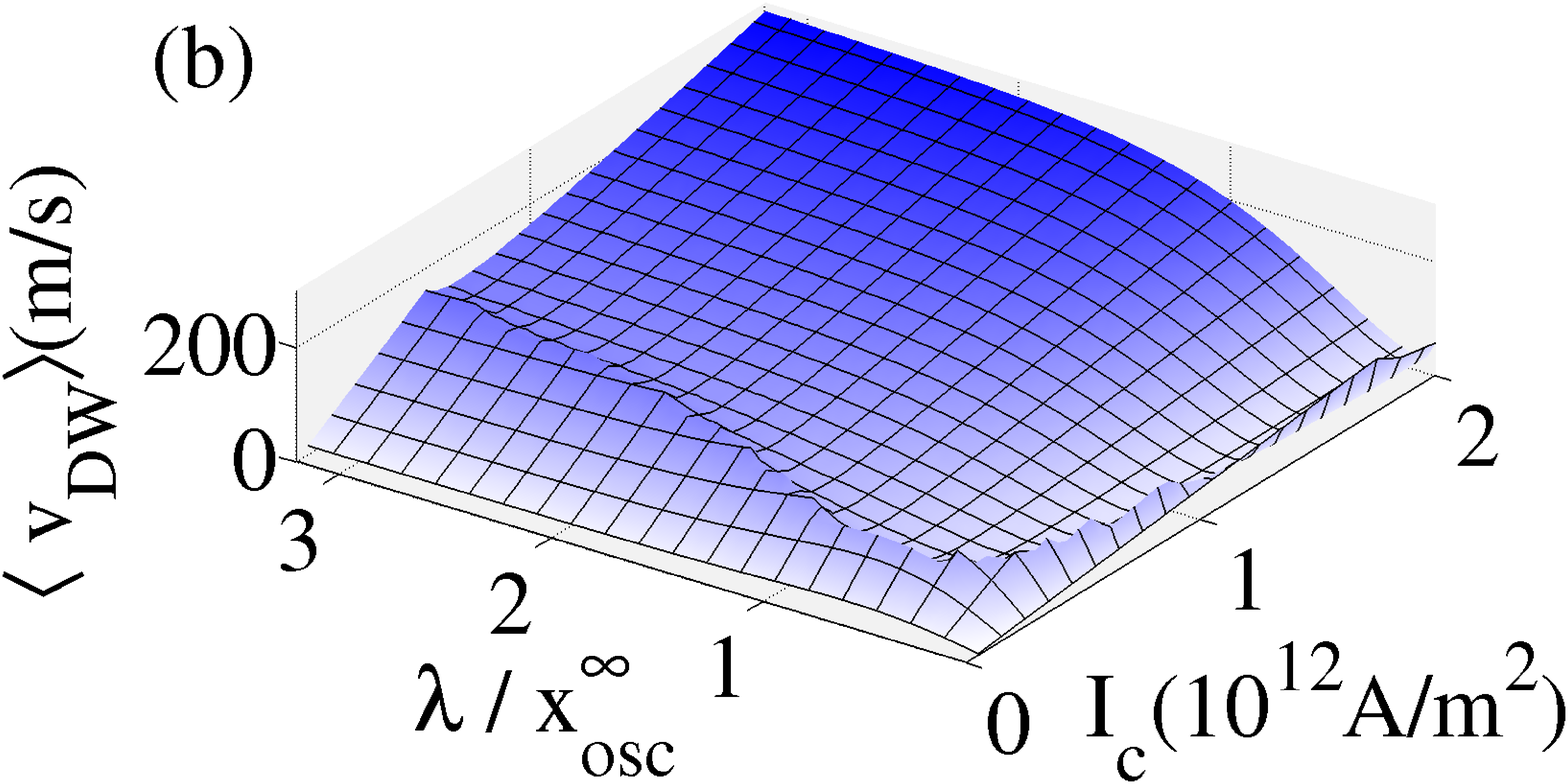}
 \includegraphics[width=.99\linewidth]{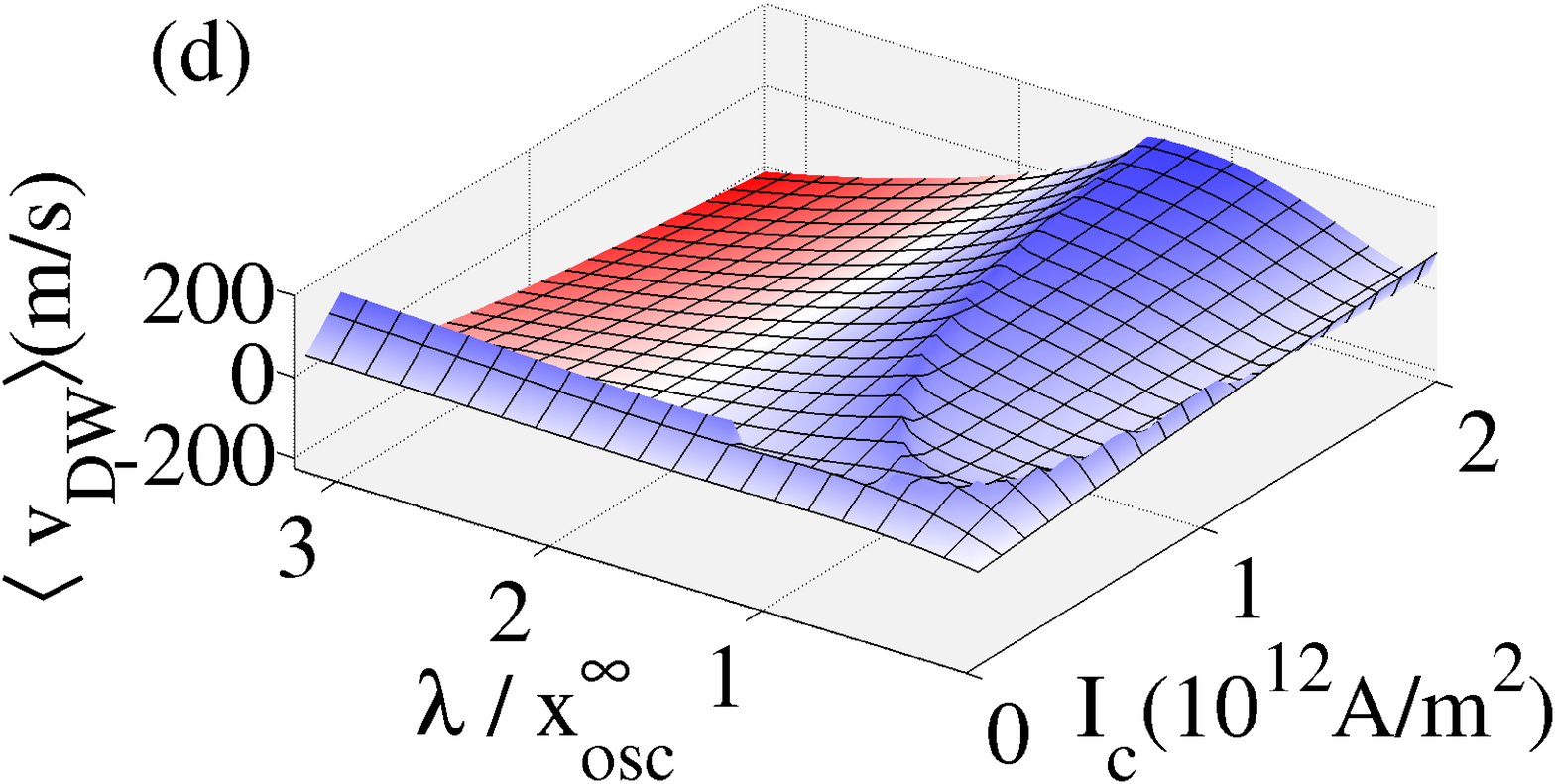}
 \includegraphics[width=.99\linewidth]{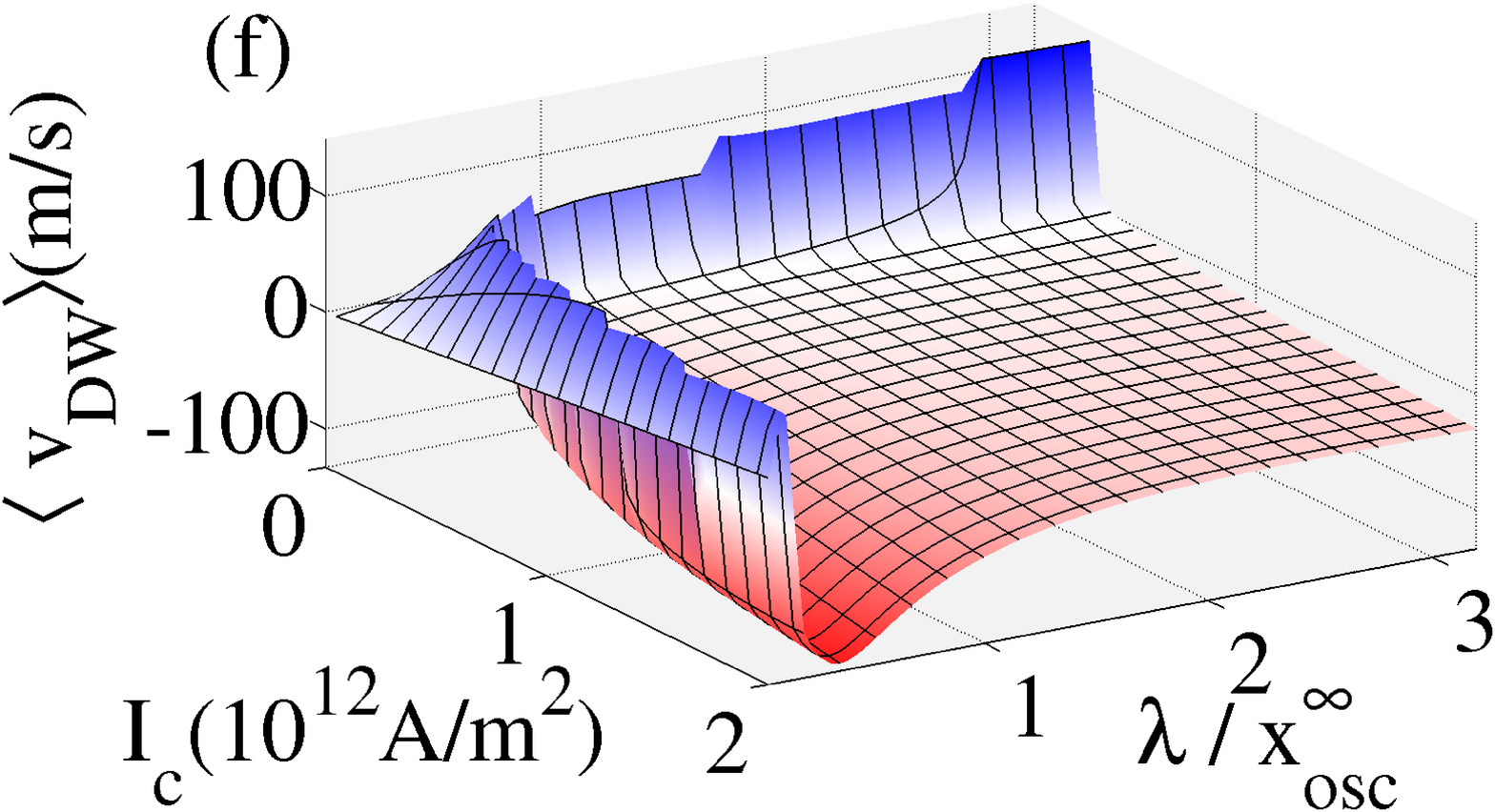}
\end{minipage}
 \caption{\label{fig::vI}(Color online) DW velocity vs DW width
$\lambda$ and applied current density $I_c$ for the cases without
relaxation ($\beta=0$) in (a,c,e), for a finite
relaxation ($\beta=0.2$) in (b,d,f) for either vanishing Rashba coupling
($\ar=0$) (a,b), finite $\ar=0.4$ (c,d), or, for  a
width dependent $\ar=0.4\frac{2}{\lambda/\xio}$ (e,f). Blue (red) colors
emphasize a positive (negative) velocity.}
\end{figure}
An interesting quantity for technological applications is the velocity 
of the DW center for a given applied current density $I_c$. 
We show the results for the combinations of $\beta=0$, $\beta=0.2$, $\ar=0$
and $\ar=0.4$ in Fig.\ \ref{fig::vI}. For broad DWs in absence of the
Rashba coupling ($\lambda\gg\xio$ and $\ar=0$), we recover the well-known
results. Without relaxation, $\beta=0$, and for small currents, there
is no DW movement. A finite velocity $\mean{v_{DW}}$ arises when the current
density exceeds a critical value. For a finite $\beta=0.2=2\alpha$, the 
Walker breakdown is observed, i.e., a strong increase of the velocity for small
current and a decrease beyond a critical $I_c$.\\
However, the picture changes significantly for steep DWs. Here, even for
$\beta=0$, a finite DW velocity arises for $0<\lambda<\xio$. It actually
resembles the Walker breakdown for $\beta>\alpha$ in the case of broad DWs. For 
$\lambda \to 0$, the DW velocity decreases again.\\
A finite Rashba field-like STT for $\ar=0.4$ mainly affects broad DWs,
since the nonadiabatic STT only can be dominant for narrow DWs. So far, we
have assumed a constant Rashba coupling independent of $\lambda$. Thus, the
Rashba field-like torque is comparable for all $\lambda$ and
every magnetic field, such as an externally applied global field, moves broader
DWs faster \cite{lucassen2009}. To remove this trivial width dependence, we
introduce a rescaled $\ar=0.4\frac{2}{\lambda/\xio}$. Indeed, the DW
velocity then becomes independent of $\lambda$ for broad DWs. However, for small
$\lambda$, a strong influence of $\mathbf H_{\text{R}}$ on $\mean{v_{DW}}$ arises. In
particular, for a finite $\beta$, even a substantial backward motion of the DW
can be generated. The motion of the DW against the current flow is induced by 
 the antidamping field \cite{stier2014,linder2014wb} $\mathbf H_{\text{R}}^{\rm anti}$
which is implicitly included in the STT. \\
\begin{figure}[tb]
\centering
 \includegraphics[width=1 \linewidth]{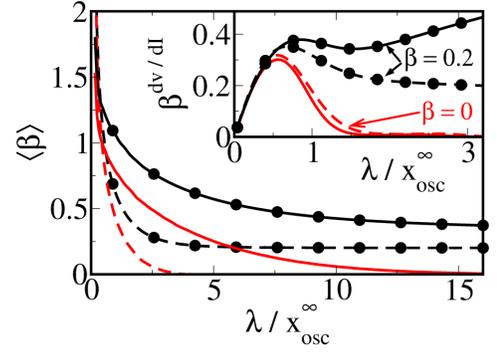}
 \caption{\label{fig::b_w}(Color online) Averaged nonadiabaticity parameter
$\mean{\beta}$ vs DW width for
$\beta=0$ (red, no symbols) and $\beta=0.2$ (black, circles). Solid lines
correspond to $\ar=0.4$ while dashed
lines refer to $\ar=0$. The inset shows $\beta^{dv/dI_c}$ obtained from the
derivative $dv/dI_c$ at small $I_c\to0$. The values in the inset are normalized such that
$\beta^{dv/dI_c}(\lambda\to\infty,\ar=0) = \beta$.}
\end{figure}
It is illuminating to discuss the case of small $\lambda$ in terms of
the nonadiabaticity parameter $\beta^*=T^{\rm{nonad}} /
T^{\rm{ad}}$ \cite{gonzalez2012}. For a nonlocal STT, $\beta^*$ 
 becomes $x$-dependent with singularities at the roots of $T^{\rm{ad}}$. Hence, 
 an averaged nonadiabaticity parameter
\begin{equation}
 \mean{\beta}=\frac{\int\beta^*(x)(\dx\fn)^2 dx}{\int (\dx\fn)^2 dx}
\end{equation}
has been introduced \cite{gonzalez2012}. In Fig. \ref{fig::b_w}, we
show $\mean{\beta}$ for small current densities. For broad DWs and
$\ar=0$, the expected result $\mean{\beta}=\beta$ is recovered. In
contrast, for steep DWs, $\lambda\lesssim \xio$, the nonadiabaticity strongly
increases. This clearly shows that the standard description of the STT fails.
What is more, for finite $\ar$, even for broad DWs, it holds that  
$\mean{\beta}\neq\beta$, because the (non)adiabatic STT also implicitly includes
the antidamping field and the standard solutions for these STTs no longer hold.\\
An alternative characterization of the nonadiabaticity results from 
observing that, for a constant STT, the DW velocity
increases with $\textrm{d}\mean{v_{DW}}/\textrm{d}I_c\propto \beta$ for small
current densities \cite{lucreview}. Thus, we define 
$\beta^{\textrm{d}v/\textrm{d}I_c}$ by the derivative of the DW 
velocity with respect to $I_c$ and normalize it to the 
conventional $\beta$ valid for large $\lambda$. The inset in Fig.\ 
\ref{fig::b_w} shows that also $\beta^{\textrm{d}v/\textrm{d}I_c}$ becomes
maximal around $\lambda\approx\xio$, but approaches zero for
$\lambda\to
0$. This further underpins the non-trivial behavior for steep
DWs which is not captured by the average $\mean{\beta}$.\\
Finally, our findings apply also to magnetic textures other than
Bloch DWs. For example, the increase of the nonadiabaticity in steep textures 
is responsible for the differences between steep vortices and broad spin waves
\cite{roessler2014,sekiguchi2012}, with an increased
$\beta^{\text{vortex}}\approx5\beta^{\text{spin-waves}}$.\\
In summary, we have introduced a general procedure to calculate the complete  
spin-transfer torque in arbitrary 1D magnetic textures. It applies to 
the entire range of steep to broad domain walls and also can include other
symmetry breaking interactions. Here, we have discussed the Rashba
spin-orbit interaction. For abrupt changes of the magnetic texture, the
STT, the DW velocity and the nonadiabaticity are 
qualitatively modified including backward motion of steep DWs against the
current. This shows that steep magnetic textures require a
fully nonadiabatic description. An extension to two-dimensional structures
such as vortices or skyrmions is straight-forward for
the case of vanishing Rashba coupling, while current flow perpendicular 
to the direct current has to be considered for finite couplings.
\begin{acknowledgments}
We acknowledge support from the DFG SFB 668 (project B16).
\end{acknowledgments}

\appendix
\section{\label{sec::app}Appendix}

In this Appendix, we present the full explicit equations for the
spin-transfer torque (STT) calculated from the
differential equation (\ref{eq::dgl}). In 
addition,
we solve this differential equation in
the limiting case of infinitely wide domain walls. Finally, we show an
additional plot of the components of the STT with
vanishing relaxation $\beta=0$.

\section{Explicit expressions of the spin-transfer torque}

Starting from the Ansatz
\begin{equation}
 \fJp=a_r\fmb+b_r\dfmb+c_r\fmb\times\dfmb\ ,\label{app::ansatz}
\end{equation}
and with $\dpp\fmb= \frac{\dpp\fn}{|\fmp|}-\frac{\fmb\cdot\dpp\fn}{|\fmp|}\fmb$, 
the STT $\mathbf T =-\dsd\fn\times\sum_r\fJp$ can be calculated. Its components 
in Eq.\ (\ref{eq::stt}) are obtained as 
\begin{align}
T_{\nu}^{\rm{ad}} =&
\frac{\dsd}{\hbar}\sum_rc_r\frac{(\fmp\cdot\fn)|\dpp\fn|}{\fmp^2|\dpp\fmb|}\times 
\begin{cases} 
      vr & \nu=x \\
      1 & \nu =t 
   \end{cases}
\nonumber\\
T_{\nu}^{\rm{non-ad}}=&-\frac{\dsd}{\hbar}\sum_r
b_r\frac{|\dpp\fn|}{|\fmp||\dpp\fmb|}\times 
\begin{cases} 
      vr & \nu=x \\
      1 & \nu =t 
   \end{cases} \nonumber\\
   H_{\text{R}}=& \frac{\dsd\alpha_{\text{R}}}{\hbar\gamma_0}\sum_r
p\left(\frac{a_r}{|\fmp|}-b_r\frac{\fmp\cdot\dpp\fn}{\fmp^2|\dpp\fmp|}\right) \, .
\end{align}

\section{Solution of the differential equation for wide domain walls}

We show how the standard solutions for the STT are recovered 
in the form of \cite{li2004}
\begin{align}
 \mathbf T=& \tilde T^{\rm ad}\dx\fn+\tilde T^{\rm{nonad}}\fn\times\dx\fn 
 -H_{\text{R}}\fn\times\fey\label{app::stdSTT}
\end{align}
as the limiting case of an infinitely wide domain wall (DW) from the 
differential equation. Here, the standard expressions \cite{li2004} $\tilde T^{\rm
ad}=I_s$,
$\tilde T^{\rm{nonad}}=-\beta I_s$ and $H_{\text{R}}=\dsd\alpha_{\text{R}}
I_s/(v\hbar)$ are known. 
To derive this result, we use a slightly modified Ansatz 
\begin{equation}
 \fJp=a_r\fmb+rv\tilde b_b\dx\fmb+rv\tilde c_r\fmb\times\dx\fmb.
\end{equation}
Here, we have not normalized the derivatives and have excluded the temporal
derivatives from the beginning. With this Ansatz, a slightly modified 
differential equation results in the form 
\begin{equation}
 vp\dx \begin{pmatrix}a_r\\ \tilde b_r\\ \tilde c_r\end{pmatrix} = A_r
\begin{pmatrix}a_r\\ \tilde b_r\\ \tilde c_r\end{pmatrix} 
 + \frac{\dsd\beta}{\hbar}\begin{pmatrix}\fmb\cdot\fJpr\\ \dx\fmb\cdot\fJpr \\
(\fmb\times\dx\fmb)\cdot\fJpr \, ,
\end{pmatrix}\label{app::dgl}
\end{equation}
with $\fJpr=a_r^{(0)}\fn$ and $\sum_r v a_r^{(0)} = I_s$. The coefficient matrix
$A_r(x,t)$ now reads
\begin{align}
 A_r = \begin{pmatrix}
-\dsd\beta/\hbar & v^2|\dx\fmb|^2 & 0\\
-1 & -\dsd\beta/\hbar -f & -\dsd|\fmp|/\hbar-g\\
0 & \dsd|\fmp|/\hbar+g & -\dsd\beta/\hbar-f
            \end{pmatrix}\label{app::coeff}
\end{align}
with $f=rv(\dx\fmb\cdot\dx^2\fmb)/(\dx\fmb)^2$ and
$g=rv[\dx\fmb\cdot(\fmb\times\dx^2\fmb)]/(\dx\fmb)^2$. Since 
$\dx\fmp\propto\dx\fn\propto 1/\lambda$, we can neglect all derivatives in Eq.\ 
(\ref{app::dgl}) in the limit of a broad DW
with $\lambda\to\infty$. When we further assume constant coefficients $a_r, \tilde b_r$
and $ \tilde c_r$ and set $|\alpha_{\text{R}}|\ll 1$, Eq.\ (\ref{app::dgl}) reduces to a 
purely algebraic system of equations
\begin{align}
0=& a_r -a_r^{(0)}, \nonumber \\
0=&-a_r -\frac{\dsd\beta}{\hbar}  \tilde b_r - \frac{\dsd}{\hbar}  \tilde c_r, \nonumber\\
0=&\frac{\dsd}{\hbar}  \tilde b_r - \frac{\dsd\beta}{\hbar}  \tilde c_r.
\end{align}
The solutions are $a_r=a_r^{(0)}$, $ \tilde b_r=-a_r^{(0)}\beta\hbar/[\dsd(1+\beta^2)]$
and
$ \tilde c_r=-a_r^{(0)}\hbar/[\dsd(1+\beta^2)]$. Thus, we get the STT $\mathbf T
=-\dsd\fn\times\sum_r\fJp$ in the final form of 
\begin{align}
 \mathbf T \stackrel{|\alpha_{\text{R}}^2|,\beta^2\ll 1}{=}&
 I_s\dx\fn  -\beta
I_s\fn\times\dx\fn 
 -\frac{\dsd\alpha_{\text{R}} I_s}{v\hbar}\fn\times\fey\nonumber. 
\label{app::explstdSTT}\\
\end{align}
In fact, these are the  standard solutions for the (non)adiabatic STT and the
Rashba field-like torque.\\
\begin{figure}[tb]
 \includegraphics[width=.8\linewidth]{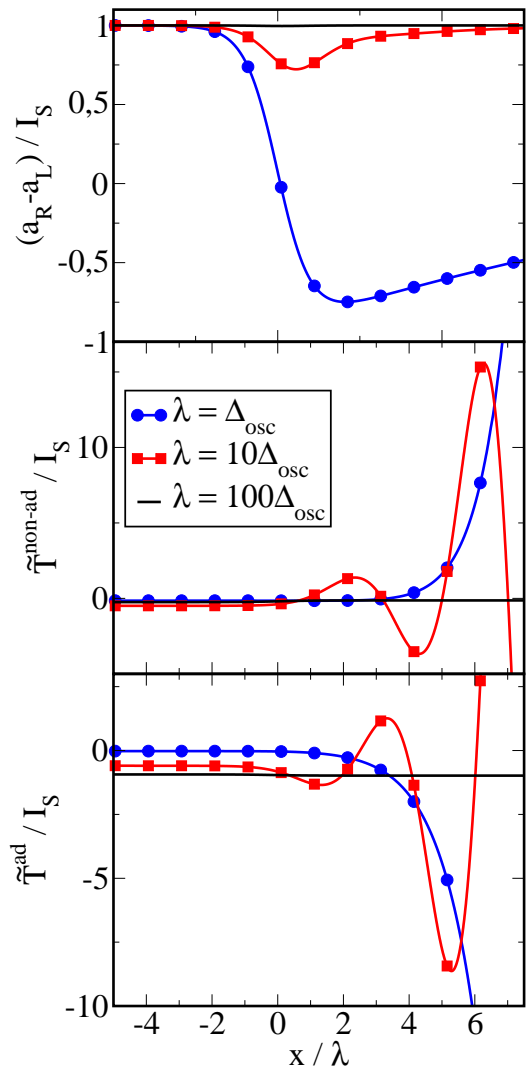}
 \caption{\label{fig::stdcomp}(Color online) Components of the STT $\mathbf
T = \tilde T^{\rm{non-ad}}\dx\fn+\tilde T^{\rm{ad}}\fn\times\dx\fn$ vs. reduced distance
from the DW center for three different
DW widths. In addition, the magnitude of the spin current density in $\fn$-direction,
$a_R-a_L$, is shown. Note that the
components are defined according to the non-normalized derivatives of $\fn$. Parameters 
are
$\ar=0$ and $\beta=0.2$.}
\end{figure}
It can be easily recognized how much the STT at small DW widths differs from the ``broad
DW limit'' in the definition of Eq. (\ref{app::stdSTT}) from numerical calculations. In
Fig. \ref{fig::stdcomp} we show the STT for several DW widths -- for simplicity for a
vanishing $\ar$. While for $\lambda=100\Delta_{\rm osc}$ the results of Eq.
(\ref{app::explstdSTT}) are recovered, the components for smaller DW widths actually
differ at and away from the DW center. This happens when the DWs are too steep to allow
the spin current density to remain aligned mainly parallel to the magnetization $\fn$, as
it can be seen in Fig. (\ref{fig::stdcomp}) from the component in $\fn$ direction. Since
the electron spin can change its orientation due to the $sd$ interaction over a length
scale of the order $x_{\rm{osc}}^{\infty}=\hbar v / \dsd$ it may be almost unchanged for
$\lambda\ll x_{\rm{osc}}^{\infty}$. As the magnetic domains change in this length
scale from $\fn(x<x_{\rm DW})\to-\fn(x>x_{\rm DW})$ it means an anti-parallel aligned
electron spin. In fact, for a perfect anti-parallel alignment and $\beta=0$ the electron
spin $\fs$ would never change back to a parallel alignment as it would provide no finite
torque $\mathbf T= -\dsd\fn\times\fs$. With a finite $\beta$ the electron spin eventually
relaxes back to the parallel alignment, but for typical $\beta \ll 1$ on an even larger
length scale $x_{\rm damp} =x_{\rm{osc}}^{\infty}/\beta$.\\
From the results above we find that a STT description containing a constant spin current
density exceeds is reliability for steep magnetic textures.

\section{Components of the spin-transfer torque without relaxation}

\begin{figure}[tb]
 \includegraphics[width=.8\linewidth]{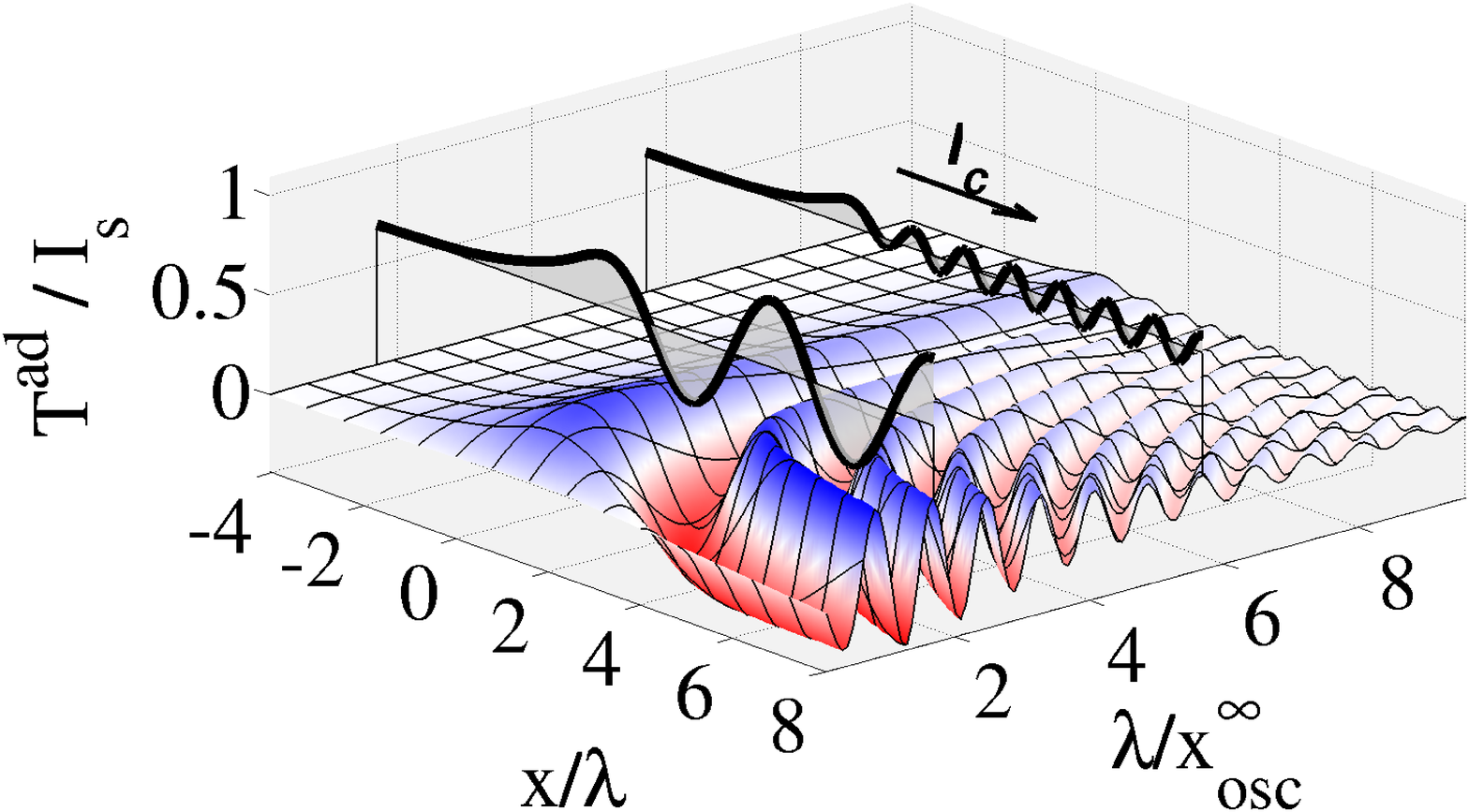}
 \includegraphics[width=.8\linewidth]{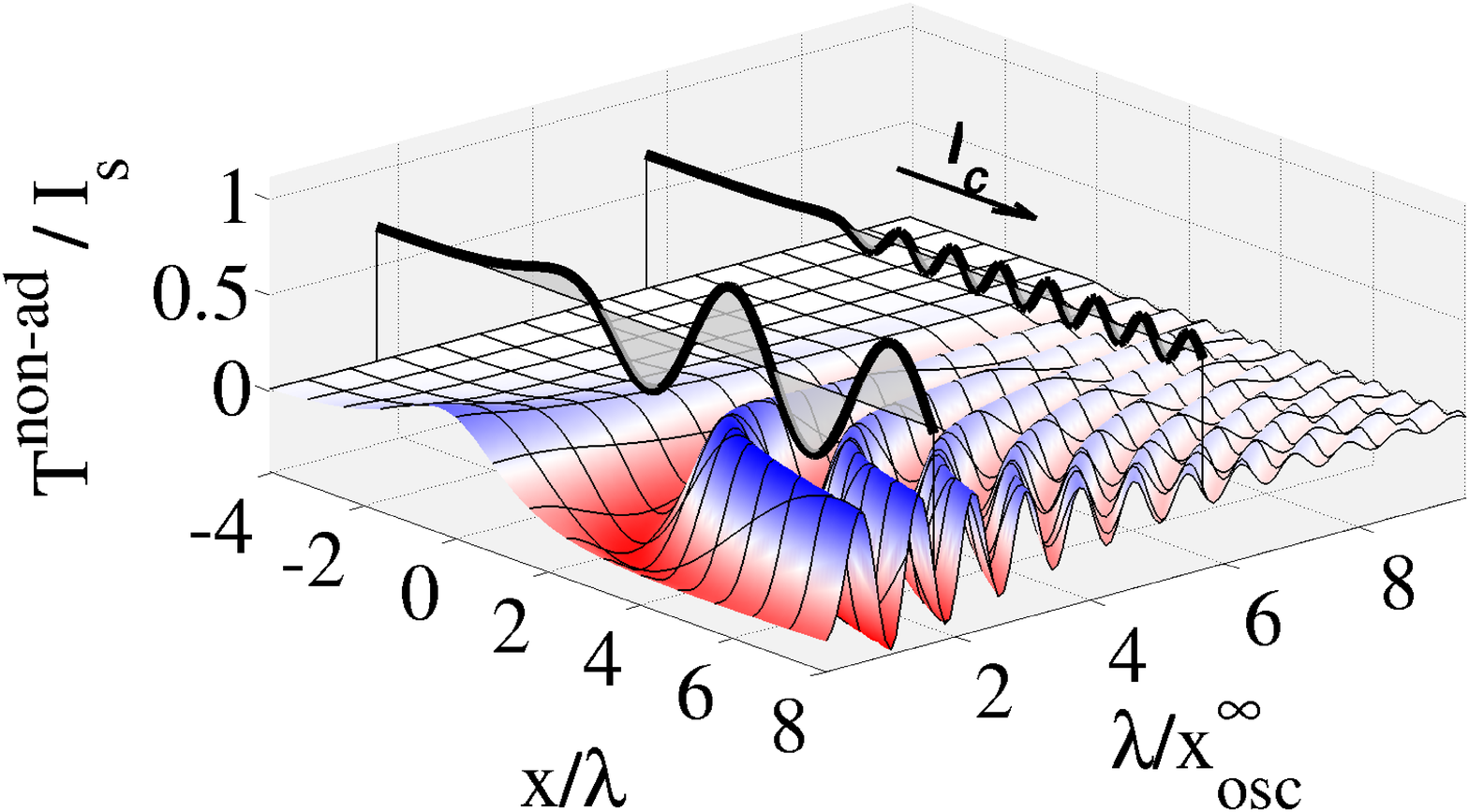}
 \includegraphics[width=.8\linewidth]{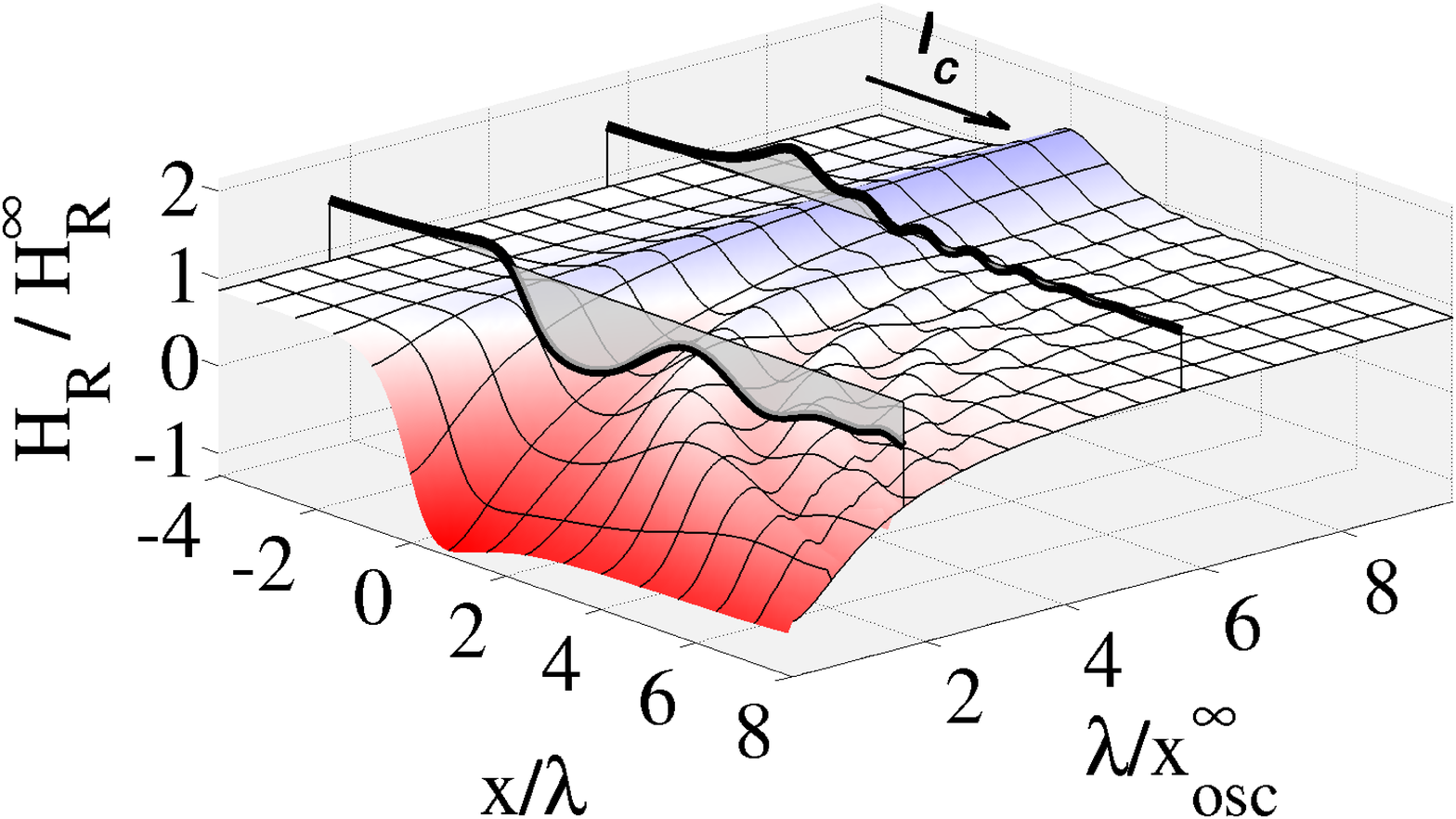}
 \caption{\label{fig::component_b0}(Color online) Spatial dependence of the
(non)adiabatic STT and the Rashba field vs 
the reduced DW width $\lambda/x_{\rm{osc}}^{\infty}$ for $\beta=0$. 
Blue (red) values indicate positive
(negative) deviations from the standard
solutions. In addition, the spatial dependence at two DW distinct widths
$\lambda/x_{\rm{osc}}^{\infty}=1.7,5.7$ is highlighted
(shifted thick black curves). No damping of the STT's oscillations occurs.  
We have set $\alpha_{\text{R}}=0.4$. }
\end{figure}
In the absence of any relaxation, i.e., $\beta=0$, neither the adiabatic nor the
nonadiabatic STT approaches the standard stationary 
solutions after the current has crossed the domain wall center $x_{DW}$. Deviations
from a non-parallel alignment are not damped and may exist over a very long distance. This
can be seen in Fig.\ 
\ref{fig::component_b0}, where we plotted the STT components according to the definition
in Eq. (\ref{eq::stt}) as it allows for more covenient physical interpretation.
Actually, oscillations with a period of $2\pi v \hbar
/\dsd $ are sustained for all $x > x_{DW}$ for both the adiabatic and the nonadiabatic
torque. This, of course, also means a finite torque far away from the DW, even though,
due to the oscillatory nature, the averaged force on the magnetic moments may vanish.\\ 
It is interesting to realize that even without a finite $\beta$, 
a strong nonadiabatic STT arises for steep magnetic textures, i.e., 
small $\lambda$. This is connected to the
non-alignment of the itinerant electron spins and local moments 
at the DW and after passing the steep magnetic structure (cf.\ Fig.\ 1).

%

\end{document}